\documentclass[a4paper]{article}

\usepackage{INTERSPEECH2021}

\title{Knowledge Distillation for Singing Voice Detection}
\name{Soumava Paul, Gurunath Reddy M, K Sreenivasa Rao and Partha Pratim Das}
\address{
  Indian Institute of Technology, Kharagpur
}
\email{soumavap@iitkgp.ac.in, \{mgurunathreddy, ksrao\}@sit.iitkgp.ernet.in, ppd@cse.iitkgp.ac.in}

\begin{document}

\maketitle
\begin{abstract}
  Singing Voice Detection (SVD) has been an active area of research in music information retrieval (MIR). Currently, two deep neural network-based methods, one based on CNN and the other on RNN, exist in literature that learn optimized features for the voice detection (VD) task and achieve state-of-the-art performance on common datasets. Both these models have a huge number of parameters (1.4M for CNN and 65.7K for RNN) and hence not suitable for deployment on devices like smartphones or embedded sensors with limited capacity in terms of memory and computation power. The most popular method to address this issue is known as knowledge distillation in deep learning literature (in addition to model compression) where a large pre-trained network known as the teacher is used to train a smaller student network. Given the wide applications of SVD in music information retrieval, to the best of our knowledge, model compression for practical deployment has not yet been explored. In this paper, efforts have been made to investigate this issue using both conventional as well as ensemble knowledge distillation techniques. 
  
\end{abstract}

\section{Introduction \& Related Work}
\label{sec:Intro}

Singing voice detection is a binary classification problem in music information retrieval, where the task is to identify singing voice in an audio segment of duration 100 to 200 ms. Efficient voice detection systems can also aid in other MIR tasks such as melody extraction~\cite{Hsu2009SingingPE, reddy2016predominant} or artist recognition~\cite{berenzweig2002using}. Early voice detection approaches~\cite{lehner2013towards, lehner2014reduction, ramona2008vocal} usually relied on complex hand-engineered audio features, which have now gone out of favour with the advent of deep neural networks. Some examples of such features include MFCCs (Mel-Frequency Cepstral Coefficients), PLPs (Perceptual Linear Predictive Coefficients) and LFPCs (Log Frequency Power Coefficients). These were used as inputs to classification systems like Support Vector Machines, Hidden Markov Models, Random Forests or Artificial Neural Networks. Two of the most popular and recent DNN based approaches~\cite{leglaive2015singing} and~\cite{schluter2015exploring} showed that neural networks have the capacity to learn complex features relevant to a particular task such as SVD, from very low-level audio representations such as STFT or Melspectrograms.~\cite{lee2018revisiting} performed evaluation of these two approaches under a common benchmark protocol so that pre-processing procedures do not give any of the two models an undue advantage in terms of performance. 

While these models have considerably improved voice detection performance as well as eliminated reliance on hand-crafted features, little attention has been paid to the optimal structure of these models - whether we actually need the huge memory and computation power needed to train these models. Also, from a deployment point of view, these networks are all non-optimal owing to their sufficiently large inference times. In this paper, we try to address this issue using principles of knowledge distillation~\cite{hinton2015distilling}. Our experiments show that the optimal CNN-based model is actually more than 250 times smaller than the model used by~\cite{schluter2015exploring}, and the optimal RNN-based model is at least 2.5 times smaller than the one used by~\cite{leglaive2015singing}. Moreover, we show that, with knowledge distillation on smaller student models, it is possible to obtain accuracies substantially higher than the current state-of-the-art.

\section{Proposed Methods}
\label{sec:prop}

\subsection{Basics of Knowledge Distillation}
\label{subsec:kd_intro}
The basic idea behind knowledge distillation is to have the student network trained with not only the information provided by true labels (also called hard targets) but also by using soft targets produced by the teacher network (also referred to as cumbersome model) as a regularizer. This helps the student network learn to mimic the teacher's behaviour. In accordance with \cite{hinton2015distilling}, a scaling hyperparameter referred to as temperature is used to regulate the softness of the targets from the teacher network. 

The generalized formula for softmax is given by:
\begin{equation}
p_{i}=\frac{\exp \left(s_{i} / \tau\right)}{\sum_{j} \exp \left(s_{j} / \tau\right)}
\tag{1}
\end{equation}
where $s_i$  and $p_i$ are the logit produced for the $i^{th}$ class and the corresponding class probability respectively and $\tau$ is the softness regulating temperature. \\

A KL-divergence loss is taken between the softmax probabilities of the teacher and the student, represented as $q$ and $p$ respectively, raised to temperature $\tau$ (between 2-20 in our experiments) and is given by:
\begin{equation}
{L}_{K D}=\tau^{2} K L D\left(q, p\right)
\tag{2}
\end{equation}
The combined loss is given by:. 
\begin{equation}
{L}_{\text {Total}}=(1-\lambda) {L}_{C E}+\lambda {L}_{K D} 
\tag{3}
\end{equation}
where ${L}_{C E}$ is the cross-entropy loss between the student's softmax probabilities (at $\tau=1$) and the correct labels (hard targets) and $\lambda$ is a second hyperparameter controlling the trade-off between the 2 losses. Generally, better results are obtained by keeping $\lambda$ close to 1, i.e., weight on ${L}_{K D}$ higher.


\subsection{Knowledge Distillation with Schluter CNN}
\label{subsec:kd_scnn} 

Before starting the discussion specific to knowledge distillation, we first describe the model architecture and input features in Schluter CNN (Teacher model) \cite{schluter2015exploring}:

\subsubsection{Input Features}

\cite{schluter2015exploring} uses melspectrograms (80-D) of 115 consecutive frames (1.6 seconds) of an audio signal as input feature. The label of the central frame is considered as the label of a particular sample. The resulting input dimension is 80x115. The Mel bank is normalized to have zero mean and unit variance over the training data.

\subsubsection{Model Architecture}
\label{subsubsec:scnn_arch}

The Schluter CNN employs three types of feedforward neural network layers: 2D convolutional layers with 3x3 kernels and Leaky ReLU activations~\cite{xu2015empirical}, maxpooling layers with 3x3 kernels and dense layers. To describe the architecture, we use the following shorthand notations: 'ConvX' denotes a conv layer with X channels and Leaky ReLU activation, 'Max' stands for maxpool, and 'DenseX' denotes a dense layer with X neurons. The architecture can then be described as Conv64-Conv32-Max-Conv128-Conv64-Max-Dense256-Dense64-Dense2. Additionally, there are two dropout layers with dropout probability 0.2 in between the final 3 dense layers. Note that the original model was trained with a sigmoid output activation function and binary cross entropy loss where appropriate thresholds were used on the output probability (between 0 and 1) to infer the presence or absence of a voice label. However, since the main objective of this paper is knowledge distillation, we needed a softmax activation on the output. Hence, in our modified model, the output dimension is 2 with a softmax activation and the model is trained with categorical cross-entropy loss. In Table \ref{tab:scnn_repr}, we compare the performance of our base model with the benchmarks by \cite{lee2018revisiting} and show that the two sets of accuracies are almost identical. This justifies the above mentioned modification to Schluter CNN.

\subsubsection{Building Student Networks from Base Model}

\begin{figure} [!htbp]
\centering
\includegraphics[scale=0.5]{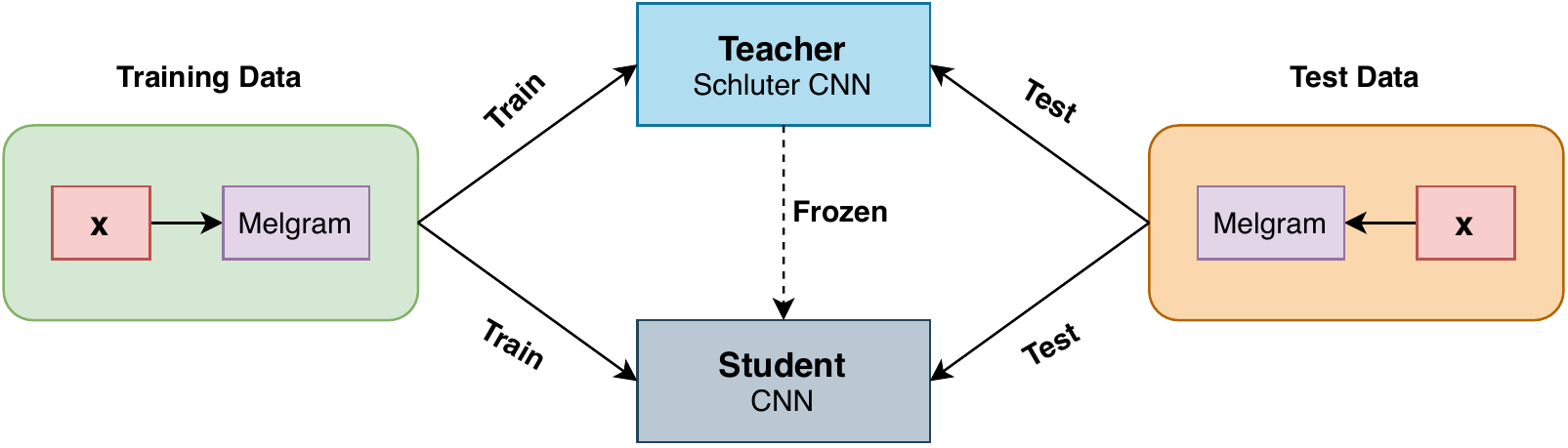}
\caption{Distilling Schluter CNN to student CNN}
\label{fig:schulter}
\vspace{-0.2cm}
\end{figure}

The base model (Schluter CNN) has a total of 14,08,290 ($\sim$ {1.4M}) parameters. To build student models, we focus on reducing the number of channels per layer (except maxpool of course). The lowest number of channels in any particular layer in the architecture mentioned above is 32. We define a variable "filter scale" (abbreviated as FS) that can have any of the values in \{2, 4, 8, 16, 32\}. As the name suggests, we divide the number of channels in each layer by FS to reduce the overall number of model parameters. Higher the value of FS, lower the number of parameters. We denote each resulting student network as FSX. Student networks trained with distillation are denoted as KD-FSX. These notations are later used in Section \ref{subsec:scnn_expt}. As shown in Table \ref{tab:scnn_derivatives}, the highest reduction achieved with student networks in terms of parameters is nearly 1000x with FS32. The process of knowledge distillation from teacher CNN model to student CNN model is shown in Fig.~\ref{fig:schulter}. As shown in figure, teacher model is the pre-trained Schluter CNN. During distillation, teacher model parameters are frozen and the knowledge of the teacher model is transferred to a small student CNN model with reduced parameters.

\begin{table}[!htbp]
\centering
\caption{Number of parameters in Student Models}
\label{tab:scnn_derivatives}
\begin{tabular}{llll}
\hline
Model & Parameters \\\hline
FS2 & 3,52,402 \\
FS4 & 88,266 \\
FS8 & 22,150 \\
FS16 & 5,580 \\
FS32 & 1417 \\\hline
\end{tabular}
\vspace{-0.5cm}
\end{table}

\begin{figure} [!htbp]
\centering
\includegraphics[scale=0.45]{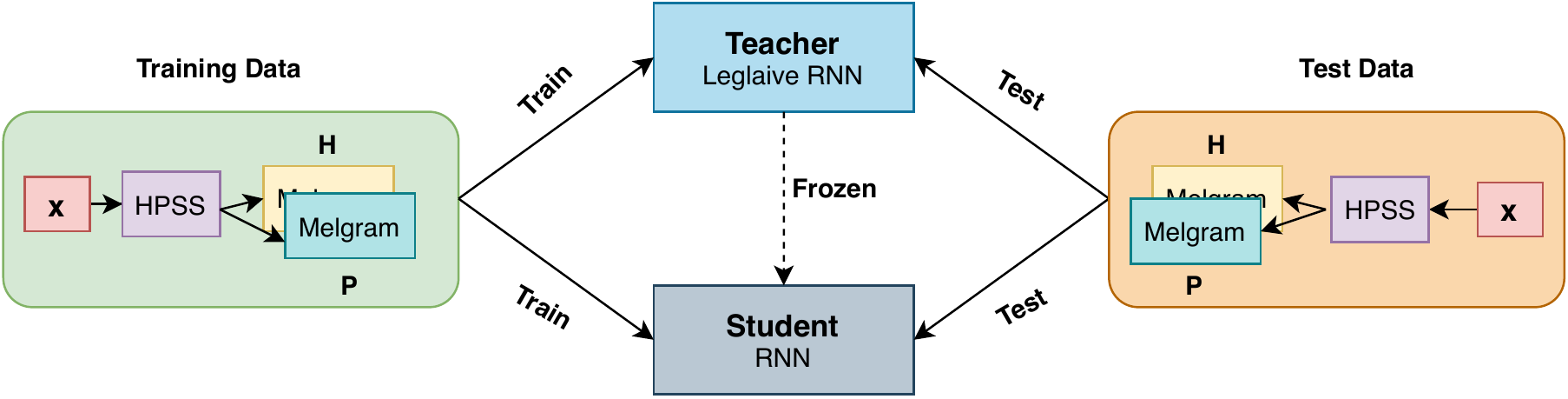}
\caption{Distilling Leglaive RNN to student RNN}
\label{fig:leglaive}
\vspace{-0.2cm}
\end{figure}

\begin{figure} [!htbp]
\centering
\includegraphics[scale=0.5]{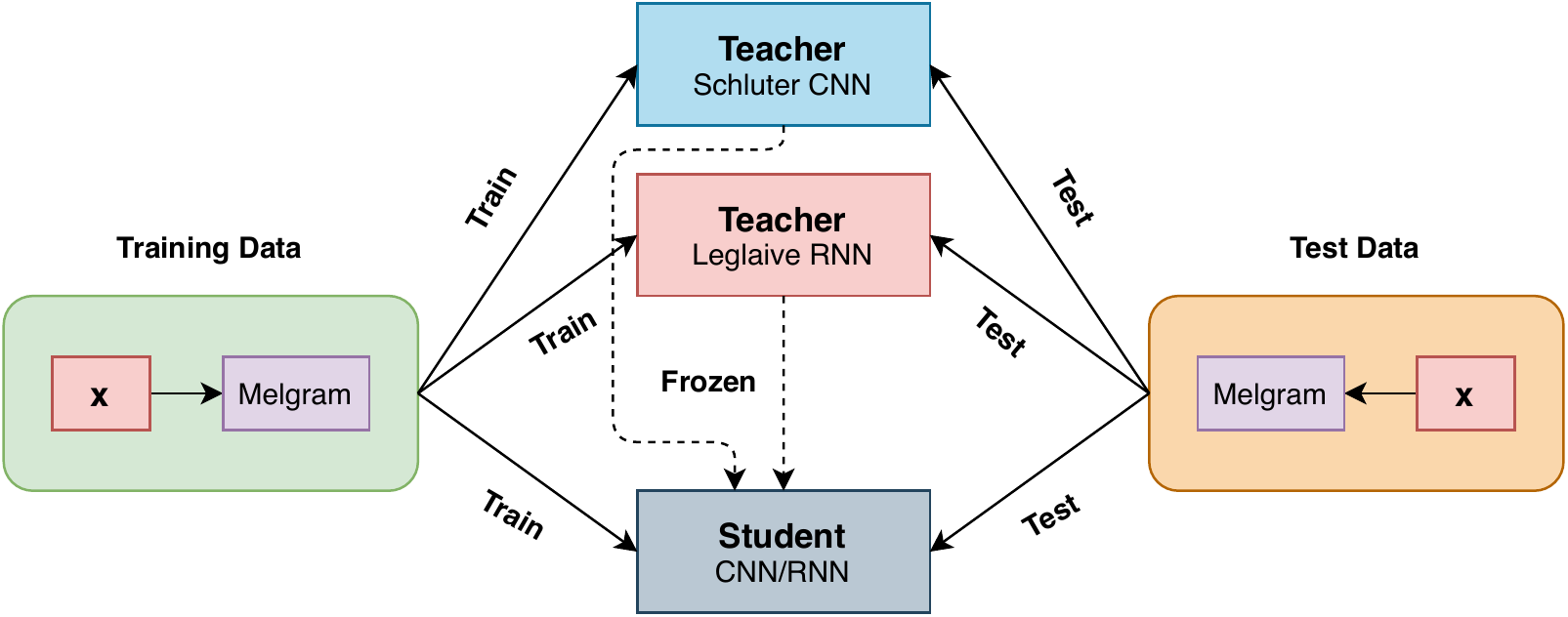}
\caption{Ensemble distillation of Schluter CNN and Leglaive RNN  to student CNN and RNN model}
\label{fig:distill}
\vspace{-0.2cm}
\end{figure}

\subsection{Knowledge Distillation with Leglaive RNN}
\label{subsec:kd_lstm}

Similar to the CNN case, we first describe the model architecture and feature inputs in Leglaive RNN (Teacher model) \cite{leglaive2015singing}:

\subsubsection{Input Features}

Leglaive RNN \cite{leglaive2015singing} first applies a double stage harmonic-percussion source separation (HPSS) \cite{ono2008separation} on the audio signal as pre-processing to extract signals relevant to the singing voice. The main idea behind HPSS is to decompose the spectrogram of the input signal into one spectrogram smooth in time direction (associated to harmonic components), and another spectrogram smooth in frequency direction (associated to percussive components). Features are then extracted from a filter bank on a Mel scale and the resulting melspectrograms of the harmonic and percussive components are concatenated as input to the Bidirectional LSTM (BiLSTM) network. This gives 80-dimensional feature vectors for 218 frames (3.5 seconds) per sample. Additionally and similar to Schluter CNN, the mel bands are normalized to have zero mean and unit standard deviation over the train set. However, unlike Schluter CNN, Leglaive RNN infers the label of each of the 218 time frames.

\subsubsection{Model Architecture}
\label{subsubsec:lstm_arch}

The model is made up of 3 layers of stacked BiLSTM units of sizes 30, 20, and 40. Finally, this is followed by a shared dense layer (also known as time distributed dense) to yield predictions for the 218 output time frames (3.5 seconds). Similar to \ref{subsubsec:scnn_arch}, the original model was trained with a sigmoid output (output dimension = 1) for each of the 218 frames and binary cross entropy loss. Keeping in mind our main objective of experimentation with distilled student models, we re-implement Leglaive RNN with 2 output dimensions with softmax activation, for each frame. Comparison with \cite{lee2018revisiting}'s implementation is shown in Table \ref{tab:lrnn_repr}. Almost identical numbers justify these modifications to the teacher network.

\subsubsection{Building Student Network from Base Model}
\label{subsubsec:lstm_params}

Leglaive RNN has a total of 65,682 ($\sim$ 65.7K) parameters. Reducing only the number of LSTM units would not have too much of an effect in terms of parameter count reduction. So, we completely remove the last 2 BiLSTM layers and that gives close to 2.5x reduction in size for the student model with a total of 26,762 ($\sim$ 26.8K) parameters. For future reference, the teacher is abbreviated as LRNN (Leglaive RNN) and the student as SRNN (small RNN). Student network trained with distillation is denoted as KD-SRNN. Knowledge distillation from teacher RNN to student RNN model is shown in Fig.~\ref{fig:leglaive}. As shown in figure, teacher model is the pre-trained Leglaive RNN. During distillation, teacher model parameters are frozen and the knowledge of the teacher model is transferred to a small student RNN model with reduced parameters.

\vspace{-0.2cm}
\subsection{Ensemble Knowledge Distillation}
\label{subsec:enkd_theory}

In this section, we propose to train both CNN and RNN based student networks using the two teachers Schluter CNN and Leglaive RNN. For this, we use the  input features representation used for SVD in~\cite{schluter2015exploring} without any data augmentation. 
However, for this feature representation, we do not have any prior RNN based models in literature. To get around this problem, we train the Leglaive RNN architecture on this data, the only difference being, the RNN now predicts labels of the central time frame as in the case of Schluter CNN. The parameter count remains unchanged from \ref{subsubsec:lstm_params}. We first benchmark the performance of LRNN and SRNN on the new data in Table \ref{tab:rnn_on_scnn_feat}. The input dimension of Schluter CNN is same as before (80x115); that of Leglaive RNN is 115x80 (time dimension comes first in RNNs). Similar to the proposition in \cite{hinton2015distilling}, we combine the predictions of the teachers in the ensemble (here 2) by taking the arithmetic or geometric mean of their predictions as soft targets. The CNN based students are denoted as ENKD-FSX and the RNN based one as ENKD-SRNN. These notations are reused in Section \ref{subsec:enkd}. As shown in Table \ref{tab:enkd_all}, we achieve state-of-the-art performance on the Jamendo dataset with this formulation. The ensemble knowledge from two teacher models to a student model is shown in Fig.~\ref{fig:distill}. As shown in figure, during knowledge distillation, the teacher (Schluter CNN and Leglaive RNN) model parameters are frozen and the knowledge of both teacher models are transferred to student CNN and RNN models.     

\section{Experiments}
\label{sec:all_expts}

\subsection{Dataset and Protocol}

For our experiments, we use the Jamendo dataset, a publicly available dataset with singing voice annotations, that was also used by \cite{leglaive2015singing} and \cite{schluter2015exploring}. It contains 93 copyright-free songs, collected and annotated by \cite{ramona2008vocal}. For comparison to existing results \cite{lee2018revisiting}, we follow the official split of 61 files for training and 16 files each for validation and testing. We use the validation set for hyperparameter tuning and report test results corresponding to the highest validation accuracy. In accordance with \cite{lee2018revisiting}, for test set, we report accuracy, precision, recall, F-measure, False Positive Rate (FPR) and False Negative Rate (FNR). Our source code is built on top of a public repository by \cite{lee2018revisiting} available here - https://github.com/kyungyunlee/ismir2018-revisiting-svd.

\subsection{Reproducing Results of Schluter CNN}
\label{subsec:scnn_repr}

Here we compare the results of our implementation (output dimension 2 and trained by categorical cross-entropy loss) with the implementation by \cite{lee2018revisiting}. Note that, similar to \cite{lee2018revisiting}, we also do not apply data augmentation for a fair comparison with \cite{leglaive2015singing}. As shown in Table \ref{tab:scnn_repr}, our numbers only differ by small magnitudes.

\begin{table}[!htbp]
\center
\scriptsize
\caption{Comparison of our Schluter CNN implementation with \cite{lee2018revisiting}'s implementation}
\label{tab:scnn_repr}
\renewcommand{\arraystretch}{1.2}
\begin{tabular}{|c|c|c|c|c|c|c|}
\hline
Model & Acc. & Prec. & Recall & F-Measure & FPR & FNR \\ 
\hline
\cite{lee2018revisiting} & \textbf{86.8} & \textbf{83.7} & 89.1 & \textbf{86.3} & \textbf{15.1} & 10.9 \\ \hline
\textbf{Ours} & 85.4 & 81.3 & \textbf{89.3} & 85.1 & 17.9 & \textbf{10.7} \\ \hline
\end{tabular}
\vspace{-0.5cm}
\end{table}

\begin{table}[!htbp]
\center
\scriptsize
\caption{Performance of Student Derivatives from Schluter CNN}
\label{tab:scnn_student}
\renewcommand{\arraystretch}{1.2}
\begin{tabular}{|c|c|c|c|c|c|c|}
\hline
Model & Acc. & Prec. & Recall & F-Measure & FPR & FNR \\ 
\hline
FS2 & \textbf{87.0} & 81.8 & \textbf{92.8} & \textbf{86.9} & 18.0 & \textbf{7.2} \\ \hline
FS4 & 85.2 & 80.4 & 90.1 & 85.0 & 19.1 & 9.9 \\ \hline
FS8 & \textbf{87.0} & 84.0 & 89.1 & 86.5 & 14.8 & 10.9 \\ \hline
FS16 & 86.5 & 84.5 & 86.6 & 85.7 & 13.5 & 13.4 \\ \hline
FS32 & 83.7 & \textbf{85.8} & 77.9 & 81.7 & \textbf{11.2} & 22.1 \\ \hline
\end{tabular}
\vspace{-0.2cm}
\end{table}

\begin{table}[!htbp]
\center
\scriptsize
\caption{Performance of Student Models with Knowledge Distillation}
\label{tab:scnn_kd}
\renewcommand{\arraystretch}{1.2}
\begin{tabular}{|c|c|c|c|c|c|c|}
\hline
Model & Acc. & Prec. & Recall & F-Measure & FPR & FNR \\ 
\hline
KD-FS2 & 87.0 & 83.7 & 89.4 & 86.4 & 15.2 & 10.6 \\ \hline
KD-FS4 & \textbf{87.3} & \textbf{84.5} & 88.3 & \textbf{86.6} & \textbf{13.6} & 11.7 \\ \hline
KD-FS8 & 86.6 & 83.1 & 89.5 & 86.2 & 15.9 & 10.5 \\ \hline
KD-FS16 & 85.0 & 80.0 & \textbf{90.4} & 84.9 & 19.7 & \textbf{9.6} \\ \hline
KD-FS32 & 81.5 & 79.4 & 81.3 & 80.3 & 18.4 & 18.7 \\ \hline
\end{tabular}
\vspace{-0.2cm}
\end{table}

\begin{table}[!htbp]
\vspace{-0.3cm}
\center
\scriptsize
\caption{Comparison of our Leglaive RNN implementation with \cite{lee2018revisiting}}
\label{tab:lrnn_repr}
\renewcommand{\arraystretch}{1.2}
\begin{tabular}{|c|c|c|c|c|c|c|}
\hline
Model & Acc. & Prec. & Recall & F-Measure & FPR & FNR \\ 
\hline
\cite{lee2018revisiting} & 87.5 & \textbf{86.1} & 87.2 & 86.6 & \textbf{12.2} & 12.8 \\ \hline
LRNN (\textbf{Ours}) & \textbf{88.2} & 85.7 & \textbf{89.7} & \textbf{87.8} & 13.2 & \textbf{10.3} \\ \hline
\end{tabular}
\vspace{-0.5cm}
\end{table}

\begin{table}[!htbp]
\center
\scriptsize
\caption{Performance Comparison of SRNN and KD-SRNN}
\label{tab:lrnn_kd}
\renewcommand{\arraystretch}{1.2}
\begin{tabular}{|c|c|c|c|c|c|c|}
\hline
Model & Acc. & Prec. & Recall & F-Measure & FPR & FNR \\ 
\hline
SRNN & 87.6 & 84.5 & 90.6 & 87.5 & 14.6 & 9.4 \\ \hline
KD-SRNN & \textbf{88.9} & \textbf{85.7} & \textbf{91.5} & \textbf{88.5} & \textbf{13.4} & \textbf{8.5} \\ \hline
\end{tabular}
\vspace{-0.5cm}
\end{table}

\begin{table}[!htbp]
\center
\scriptsize
\caption{Performance of LRNN and SRNN on data used by \cite{schluter2015exploring}}
\label{tab:rnn_on_scnn_feat}
\renewcommand{\arraystretch}{1.2}
\begin{tabular}{|c|c|c|c|c|c|c|}
\hline
Model & Acc. & Precision & Recall & F-Measure & FPR & FNR \\ 
\hline
LRNN & \textbf{82.3} & \textbf{78.2} & \textbf{85.8} & \textbf{81.8} & \textbf{20.9} & \textbf{14.2} \\ \hline
SRNN & 76.1 & 70.5 & 83.8 & 76.6 & 30.6 & 16.2 \\ \hline
\end{tabular}
\vspace{-0.2cm}
\end{table}

\begin{table*}[!htbp]
\center
\scriptsize
\caption{Performance Comparison of ENKD Variants}
\label{tab:enkd_all}
\renewcommand{\arraystretch}{1.2}
\begin{tabular}{|c|c|c|c|c|c|c|c|c|c|c|c|c|}
\hline
& \multicolumn{2}{c|}{Acc.} & \multicolumn{2}{c|}{Prec.} & \multicolumn{2}{c|}{Recall} & \multicolumn{2}{c|}{F-Measure} & \multicolumn{2}{c|}{FPR} & \multicolumn{2}{c|}{FNR} 
\\ \hline
Model & \begin{tabular}[c]{@{}c@{}}AM\end{tabular} & \begin{tabular}[c]{@{}c@{}}GM\end{tabular} &  \begin{tabular}[c]{@{}c@{}}AM\end{tabular} & \begin{tabular}[c]{@{}c@{}}GM\end{tabular} &  \begin{tabular}[c]{@{}c@{}}AM\end{tabular} & \begin{tabular}[c]{@{}c@{}}GM\end{tabular} &  \begin{tabular}[c]{@{}c@{}}AM\end{tabular} & \begin{tabular}[c]{@{}c@{}}GM\end{tabular} &  \begin{tabular}[c]{@{}c@{}}AM\end{tabular} & \begin{tabular}[c]{@{}c@{}}GM\end{tabular} &  \begin{tabular}[c]{@{}c@{}}AM\end{tabular} & \begin{tabular}[c]{@{}c@{}}GM\end{tabular} 
\\ \hline
ENKD-FS2 & 87.9 & 87.7 & 86.3 & 86.5 & 87.9 & 87.1 & 87.1 & 86.8 & 12.1 & 11.8 & 12.1 & 12.9 \\ \hline
ENKD-FS4 & 86.8 & \textbf{88.4} & 84.4 & \textbf{86.8} & 88.0 & 88.4 & 86.1 & \textbf{87.6} & 14.2 & \textbf{11.7} & 12.0 & 11.6 \\ \hline
ENKD-FS8 & 87.7 & 87.1 & 86.2 & 81.9 & 87.6 & \textbf{92.8} & 86.9 & 87.0 & 12.2 & 17.8 & 12.4 & \textbf{7.2} \\ \hline
ENKD-FS16 & 86.4 & 86.5 & 84.2 & 83.4 & 87.1 & 88.1 & 85.6 & 85.8 & 14.3 & 15.0 & 12.9 & 11.9 \\ \hline
ENKD-FS32 & 84.3 & 84.2 & 82.3 & 82.8 & 84.4 & 83.3 & 83.3 & 83.1 & 15.8 & 15.0 & 15.6 & 16.7 \\ \hline
ENKD-SRNN & 84.4 & 83.1 & 81.1 & 81.5 & 86.5 & 82.5 & 83.7 & 82.0 & 17.5 & 16.3 & 13.5 & 17.5 
\\ \hline
\end{tabular}
\vspace{-0.2cm}
\end{table*}

\vspace{-0.1cm}
\subsection{Knowledge Distillation with Schluter CNN}
\label{subsec:scnn_expt}
\subsubsection{Performance of Student Networks without Distillation}
\label{subsubsec:student_base}
Here we show the performance of the student networks from Table \ref{tab:scnn_derivatives}. These are trained similar to the teacher network. As shown in Table \ref{tab:scnn_student}, with our smaller student networks, we get better results across all performance measures, with respect to both ours and the original implementation in Table \ref{tab:scnn_repr}. Interestingly enough, FS32, having 1000x lesser parameters than the teacher, succeeds in improving on 2 performance measures, Precision and FPR, over Schluter CNN. This clearly indicates, that smaller models can be equally effective in SVD and require further attention.

\subsubsection{Performance of Distilled Student Networks}
Here we present the performance of the student networks from \ref{subsubsec:student_base}, trained with distillation by the teacher network. As shown in Table \ref{tab:scnn_kd}, performance falls on almost every measure with the student models. This is expected since, as per Table \ref{tab:scnn_student}, student networks already have a lesser tendency to overfit, i.e., they actually perform better than Schluter CNN. So, additional supervision from the teacher actually ends up hurting the student network. However, as we show in Section \ref{subsec:enkd}, performance of all FSX models improve with additional supervision from an LSTM-based \cite{hochreiter1997long} teacher network.


\subsection{Reproducing Results of Leglaive RNN}
Similar to Section \ref{subsec:scnn_repr}, we compare our implementation of Leglaive RNN with that of \cite{lee2018revisiting}. Here the output dimension is 2 for the time distributed dense layer and the model is trained by categorical cross-entropy loss. As shown in Table \ref{tab:lrnn_repr}, our numbers only differ by small magnitudes.

\subsection{Knowledge Distillation with Leglaive RNN}

\subsubsection{Performance of Student Network with and without Distillation}

As shown in Table \ref{tab:lrnn_kd}, KD-SRNN comprehensively outperforms SRNN on all performance measures, showing the benefits of knowledge distillation in this context. Although SRNN performs slightly worse than LRNN on some of the measures, KD-SRNN surpasses LRNN (our implementation) on all measures and is inferior to the original implementation \cite{lee2018revisiting}, only on Precision and FPR.

\subsection{Ensemble Knowledge Distillation}
\label{subsec:enkd}

\subsubsection{Benchmarking LSTM Performance}

First we show the performance of LRNN and SRNN on the data used by Schluter CNN in Table \ref{tab:rnn_on_scnn_feat}. We see that its performance is considerably lower than Schluter CNN or any of the FSX derivatives. Despite this, including it as a teacher in the ensemble considerably improves the performance of all ENKD-FSX and ENKD-SRNN models, as shown in Table \ref{tab:enkd_all}.

\subsubsection{ENKD Models}

Here we present the performance of all CNN and RNN based student networks trained by Ensemble Knowledge Distillation shown in Table \ref{tab:enkd_all}.
AM (Arithmetic Mean) and GM (Geometric Mean) denote method by which teacher predictions were combined. With Ensemble Knowledge Distillation (ENKD), we get substantial improvements over all CNN and RNN models trained by simple knowledge distillation (KD). On accuaracy, ENKD improves KD-FSX variants by the following amounts - 0.9\% (FS2), 1.1\% (FS4), 1.1\% (FS8), 1.5\% (FS16), 2.8\% (FS32). ENKD also gives a new state-of-the-art model on the features used by \cite{schluter2015exploring}, ENKD-FS4 (mainly based on accuracy) that improves current best reported results (Table \ref{tab:scnn_repr}) by 1.6\%. This technique also proves quite beneficial for SRNN, improving its accuracy by 8.3\% to 84.4\% (higher than even LRNN). However, this accuracy is still lower than that of ENKD-FS4 by a fair amount, proving that CNN is still the better model for this kind of data.

In summary, smaller student networks built from Schluter CNN actually show lesser overfitting tendencies and hence perform better than the base model in several cases. That is why knowledge distillation experiments with the base model as a teacher, fail to get any substantial improvement over the student networks' performance. Knowledge Distillation experiments with Leglaive RNN produce new state-of-the-art results on the features used by \cite{leglaive2015singing}, with a smaller student model, SRNN. With Ensemble Knowledge Distillation (ENKD), we get substantial improvements over all CNN and RNN models trained by simple knowledge distillation (KD).

    
    

\vspace{-0.1cm}
\section{Conclusion and Future Work}
\label{sec:conc}

In this paper, we have shown that application of knowledge distillation techniques to compress the singing voice detection models can be a new direction of research in the field of Music Information Retrieval. Our experiments show that smaller models trained with distillation can achieve comparable and in some cases, even higher accuracies than current state-of-the-art models. 
With respect to knowledge distillation, another interesting experiment can be training student models on a subset of the training data and comparing their performance with student/teacher models that learn from the entire data. We also plan to explore multi-step knowledge distillation techniques in the future, like the Teacher Assistant model~\cite{mirzadeh2019improved}, where an intermediate-sized network can supposedly help bridge the knowledge gap between student and teacher better and hence aid in more efficient distillation.

\section{Acknowledgements}
The second author would like to thank Google for supporting his PhD through Google PhD Fellowship program.

\bibliographystyle{IEEEtran}
\bibliography{mybib}


\end{document}